\documentclass{ptapap}

\author{Aleksandra Leśniewska}[UAM]
\author{Magdalena Otulakowska-Hypka}[UAM]
\author{Joanna Mikołajewska}[CAMK]
\author{Patricia A. Whitelock}[SAAO, ADC]

\affil[UAM]{Astronomical Observatory Institute,
A. Mickiewicz University, Poznań, Poland}
\affil[CAMK]{N. Copernicus Astronomical Center, 
Polish Academy of Sciences, Warsaw, Poland}
\affil[SAAO]{South African Astronomical Observatory, 
South Africa}
\affil[ADC]{Department of Astronomy, University of Cape Town, Private Bag X3, Rondebosch 7701, South Africa}

\title{SY Mus -- search for physical parameters}

\begin{document}

\maketitle

\begin{abstract}

This is a preliminary analysis of the orbital parameters of the well known eclipsing symbiotic binary, SY Muscae. It is a system composed of a white dwarf (WD) and a red giant (RG), located in the southern sky. With the use of photometric data in the infrared (IR) bands and radial velocities (RV) for the RG, we determine physical parameters of the object, such as masses and radii. We use PHOEBE tools to model all the observations.

\end{abstract}

\section{Introduction}

Since Annie J. Cannon separated the new type of objects based on their spectra, symbiotic stars have become widely analyzed as binary stars. Symbiotic stars are an extremely important element in understanding the evolution of binary stars. The characteristics of these objects, their physical and orbital parameters as well as their stage of evolution allow us to refine our knowledge about these stars. We chose the symbiotic star of the southern sky, SY Muscae, well known in the literature, to improve its characteristics. With the simultaneous use of different old and new observations, we aimed at better understanding properties of this type of objects. This is a system, composed of a WD and a RG, with a high inclination which allows us to observe ellipsoidal changes in the IR related to the shape of the giant as well as eclipses.

\section{Early analysis}

We have reviewed the literature to learn the known parameters of the object before modeling. One of the best defined parameters is the orbital period, $P=624.5$~d  \citep{rutkowski}. Other parameters were not so clearly defined, so during our work we focused on determining all of them, with the aim to find masses of components, their radii, separation, and the inclination of the system. Based on the literature, we set initial values for a number of PHOEBE's parameters, namely separation between objects, $a=350 R\odot$ \citep{rutkowski}, inclination $i=88.8^{\circ}$ \citep{rutkowski}, effective temperature of RG as 3400 K \citep{mikolajewska}, effective temperature of WD  as 100 000 K \citep{skopal}, metallicity of RG as \textit{[Fe/H]} = -0.15 \citep{galan}, and systemic velocity as 13.0 km/s \citep{schmutz}. In PHOEBE we use the \textit{unconstrained binary system} model. 

We used photometry in four infrared bands from the 0.75~m telescope at SAAO, which is on the photometric system defined by \cite{carter}. The errors are less than 0.03~mag in \textit{J, H} and \textit{K}, and less than 0.05~mag in \textit{L}. Radial velocity data comes from \cite{fekel}. 

\begin{figure}
   \includegraphics[height=9cm,width=1.1\textwidth]{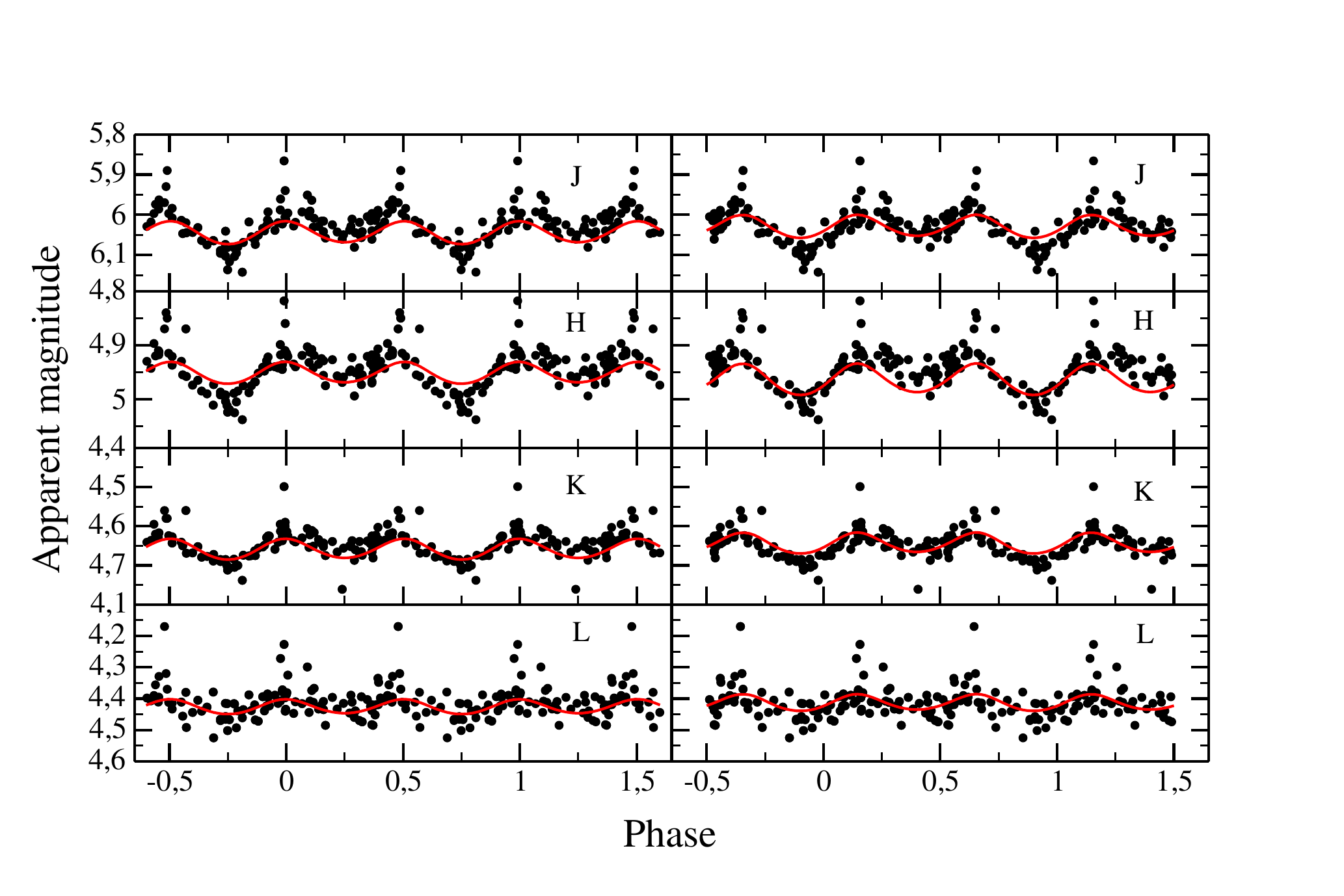}
   \caption{Comparison of two models with reversed mass ratios, i.e. $q_1 = 2.4$ (left) and $q_2 = 0.38$ (right). Data in four bands $J, H, K, L$ (black dots) and fitted curves (red lines).}
   \label{LC}
\end{figure}

Two training approaches were used to get familiar with PHOEBE and to better understand its tools. First, we set the mass ratio, $q$, as secondary over primary star, and second, as the opposite. 
Figure \ref{LC} shows four IR light curves (LC) for both the mass ratios, $q_1=2.4$ on the left, and $q_2=0.38$ on the right side.

\begin{figure}
  \centering
  \begin{minipage}{0.5\textwidth}
    \includegraphics[width=\textwidth]{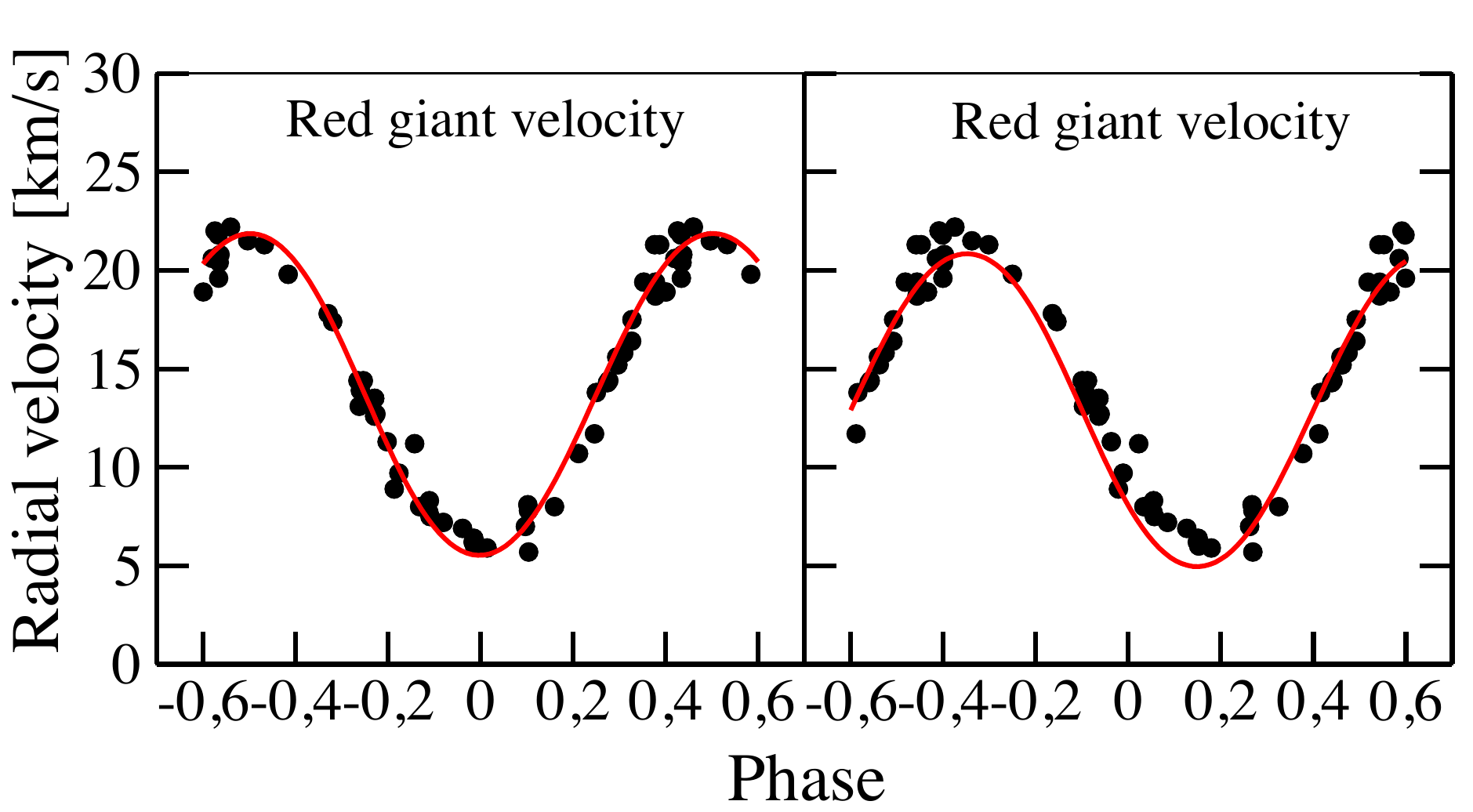}
    \caption{The RV as a function of phase for $q_1$ (left) and for $q_2$ (right). Data in the IR for red giant (black dots) and the fitted curve (red line).}
    \label{RV}
  \end{minipage}
  \quad
  \begin{minipage}{0.45\textwidth}
    \includegraphics[width=\textwidth]{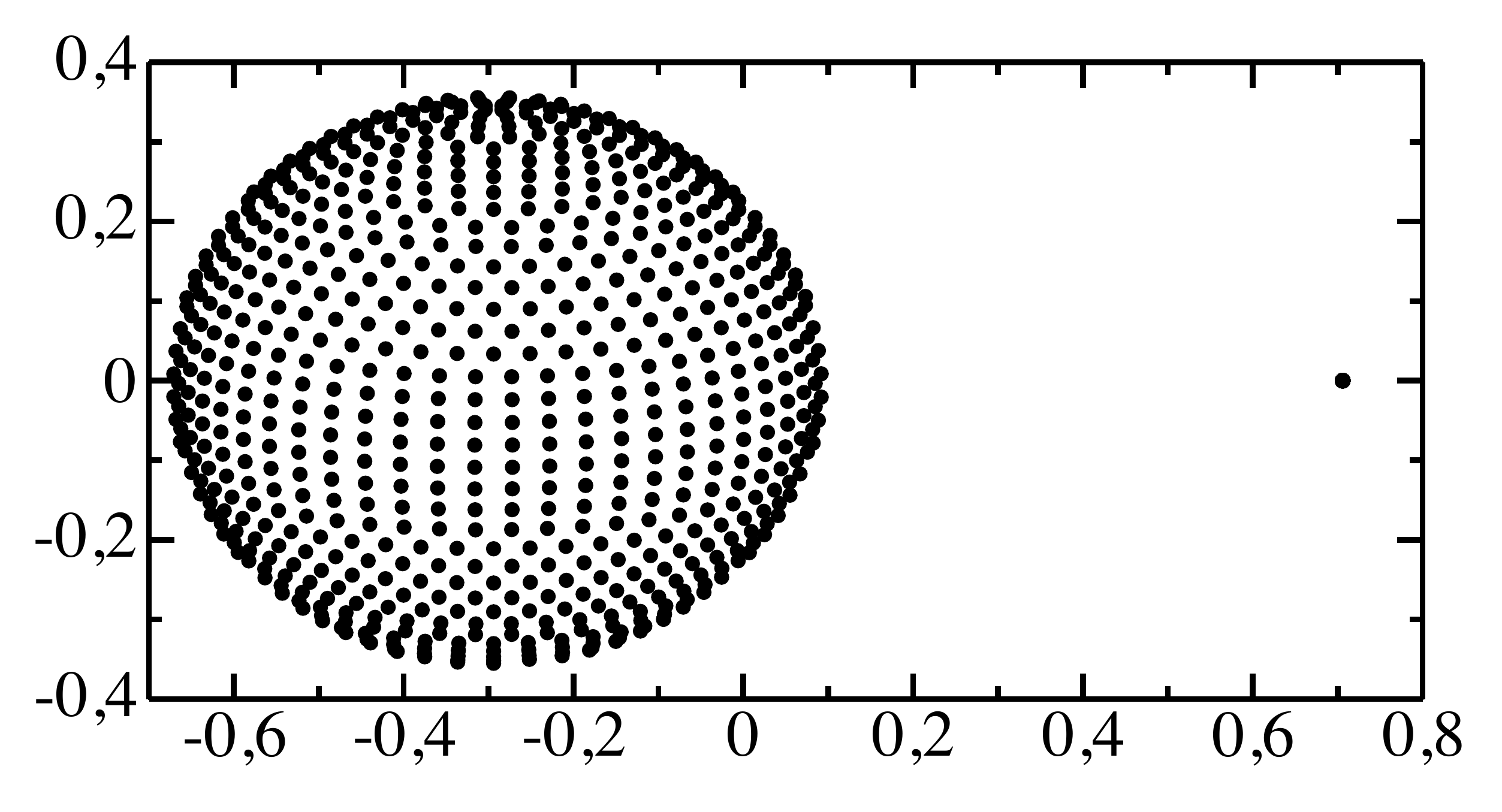}
    \caption{The system's shape showing the relative distances of components from the center of mass.
    }
    \label{system}
  \end{minipage}
\end{figure}

\section{Preliminary results}

We found approximate values of the system's basic parameters, see Table \ref{tab1}. Results from both models are similar and comparable to those in the literature. LC and RV curves in both approaches provide similar results. The fit seems to be better for RV in the case of the first model, however, for the LC, especially in the \textit{H}-band, the second model seems better.  

\begin{table}[ht]
\centering
\begin{tabular}{ccc}
\hline
 & $q_1$ = M$_g$/M$_h$ = 2.4 & $q_2$ = M$_h$/M$_g$ = 0.38 \\
\hline
i [$^{\circ}$] & 75 & 88 \\
a [R$\odot$] & 357.5 & 360  \\
M$_g$ [M$\odot$] & 1.11 & 1.17 \\
M$_h$ [M$\odot$] & 0.46 & 0.44 \\
R$_g$ [R$\odot$] & 131.22 & 143.93 \\
R$_h$ [R$\odot$] & 0.10 & 0.14 \\
V$_0$ [km/s] & 13.7 & 12.9 \\
P [d] (fixed) & 624.5 & 624.5\\

\end{tabular}
\caption{Two possible solutions of our modeling.} \label{tab1}
\end{table}

\section{Future plans}
This is a work in progress and our next stage will include additional photometric and spectroscopic data. Next, we will estimate uncertainties of the solution.

\bibliographystyle{ptapap}
\bibliography{ptapapdoc}

\end{document}